# The gruesome murder of Jamal Khashoggi : Saudi Arabia's new economy dream at risk ?

Jamal BOUOIYOUR

IRMAPE, ESC Pau Business school, France.

CATT, University of Pau, France.

E-mail: jamal.bouoiyour@univ-pau.fr

Refk SELMI

IRMAPE, ESC Pau Business school, France.

CATT, University of Pau, France.

E-mail: s.refk@yahoo.fr

**Abstract :** With the horrific Jamal Khashoggi killing, Mohammed Bin Salman's image in the international community has been damaged. This study seeks to test whether Khashoggi murder discourage businesses from investing in Saudi Arabia. We use an event-study methodology and asset pricing model to assess, at sectoral level, the dynamics of stock prices surrounding the killing of the Saudi journalist on 2 October at the kingdom's consulate in Istanbul. A series of robustness tests, including the Corrado ranking test and the non-parametric conditional distribution approach, have been conducted. We consistently show that the khashoggi killing had the most adverse impact on banks and financial services, materials, and technology. Oil and gas companies, however, were moderately or insignificantly affected. Overall, our results suggest that the crown prince's ambitious project for a Saudi Arabian economy moving beyond oil wealth are threatened as this recent event dampened foreign interest in investing in the kingdom.

1. Introduction

When Saudi Arabia's 32-year-old Mohamed Bin Salman Bin Abdulaziz Al Saud– colloquially known as MBS– ascended from deputy to Crown Prince in 21 June 2017, there was a high expectation of an exceptional change. Since assuming the role, MBS has taken great strides towards social and economic liberalization of Saudi Arabia. MBS lunched larger reforms known as 'Vision 2030'. The latter consists of diversifying the economy of the Wahhabi Kingdom, and mitigating its high reliance on oil income. More importantly, this project is built around three main themes : a vibrant society, a prosperous economy and an ambitious nation. The first theme is indispensable to fulfilling a great foundation for the economic prosperity. The main purpose is to stimulate social development in an attempt to build a productive society by offering the education that builds children's fundamental characters and fulfilling an empowering social and health care system. The second theme aims to achieve a thriving economy by improving the education system while aligning it with market needs and offering a variety of economic opportunities for businesses, both the small enterprise and the big corporations. In doing so, they want to establish the adequate investment tools helping to diversify the Saudi Arabian economy and create new job opportunities. They want also to enhance the quality of services, by privatizing some government services, improving the business climate, and spurring innovation. The third theme is built on an efficacious, transparent, accountable, and high-performing government to better encourage the private sector to take the initiative.

A successful achievement of these goals is highly conditional upon international investment attraction. The Khashoggi murder at the Saudi consulate has harmed MBS's image in the international community, and has led to mass withdrawals from the Saudi Future Investment Initiative (FII) summit, also called 'Davos of the Desert'. This would be as disturbing to Saudi policymakers. Jamal Khashoggi, 59, a Saudi journalist who was always critical of the government, entered the kingdom's consulate in Istanbul on October 2 and had not been seen since. On October 19, the Saudi government announced his death, indicating he died during an altercation in the consulate. The gruesome murder of Jamal Khashoggi has shaken the world. In fact, the Saudi journalist killing provoked heavy reactions on the international scene, especially in the business world. The FII summit aimed at attracting international investments to the country seems in crisis as Khashoggi case prompts mass withdrawals. It wasn't that long ago that Saudi crown prince, was largely regarded as someone who would escort in sharp positive changes to the Wahhabi kingdom. Today, the MBS's image is hugely threatened due to the Khashoggi's unexplained disappearance after he entered the Saudi Consulate in Istanbul. US intelligence agencies have suspected MBS ordered the killing of the journalist, contradicting the Saudi government's proclamation that the Saudi crown prince was not involved. It must be stressed that MBS was engaged in suspicious activities prior to Khashoggi murder, including the military intervention in Yemen since 2016, the Qatar diplomatic crisis (June 2017), the roundup of his rivals in November 2017, and the detention of the Lebanese prime minister Saad Hariri in December 2017, but his public image was not adversely influenced. Nevertheless, MBS's image as a reformer has been flipped on its head amid global outcry surrounding Khashoggi's death. The most immediate impacts of Khashoggi murder on MBS's big social and economic ambitions and Saudi Arabia's plans for an economy moving beyond its oil wealth were debated intensively since the disappearance of the Saudi journalist. However, there is no accordance on the relative costs of this crisis on businesses. We may get a feel of the possible responses of various sectors of the Saudi Arabia economy by examining how disaggregated stock indices reacted following the killing of the journalist in the Saudi consul-general in Istanbul on 2 October 2018.

Individual and institutional investors have long considered heightened uncertainty indispensable determinant of the stock market volatility (for instance, Arnold and Vrugt 2008 ; Bloom 2009, 2014). Heightened uncertainty over a sudden event also prompts a decline in investment, as claimed by Branden and Yook (2012) and Bøe and Jordal (2016). Accordingly, the event study methodology has been successfully carried out to examine the impact of the uncertainty surrounding different events (inter alia : terrorist attacks, Brexit, Trump's win in the 2016 US presidential elections) on stock markets. For example, Kolaric and Schiereck (2016) investigated the reactions of airline stock prices over the terrorist attacks in Paris and Brussels. Ramiah et al. (2016) and Bouoiyour and Selmi (2018 a) explored, at sectoral level, the dynamics of stock prices surrounding the announcement of the UK's EU membership referendum on 24 June 2016 (Brexit). Bouoiyour and Selmi (2018 b) tested whether BRICS stock markets were equally vulnerable to Trump's agenda using event-study methodology over a period of 120 days toward the final election result on 08 November 2016. Likewise, Pham et al. (2018) carried out the event study methodology and a variety of asset pricing models to evaluate the impacts of Trump's win in the 2016 US presidential election as well as the events happening in the run-up to election day on the US equity market. They consistently found that the events around the 2016 U.S. election may prompt diamond risk phenomenon. However, the implications of the recent Khashoggi crisis for Saudi Arabian economy remains unexplored. This study uses an event study methodology that examines the abnormal returns behaviors for several sectors of the Saudi stock market (in particular, Financials, Materials, Oil and Gas and Technology) around the date of the journalist's disappearance. The event study methodology assumes that the abnormal returns of a firm is a function of revenue minus cost. Zero abnormal returns may indicate that neither revenue nor cost changes as a consequence of Khashoggi killing, or it is a protected industry. Otherwise, positive and negative abnormal returns mean favourable and unfavourable impacts, respectively. The event study methodology has been largely criticized because of the non-normality of the abnormal returns (ARs) distribution and the effects of stock market integration. It is widely acknowledged that ARs are not normally distributed and have a tendency to display high kurtosis and positive skewness, which may have a significant impact on the parametric t-statistics. To address this criticism, this study follows Ramiah et al. (2016) and Pham et al. (2018) methodology. Specifically, we conduct a Corrado (1989) non-parametric ranking test and the non-parametric conditional distribution approach developed by Chesney et al. (2011) while controlling for stock market integration and spillover effects. Our findings indicate that, with the exception of Oil and Gas industry, all the rest of sectors were harmfully affected by Jamal Khashoggi's disappearance.

The remainder of the study is organized as follows. Section 2 presents the major obstacles facing the Saudi crown prince's vision 2030. Section 3 describes the methodology and presents the data. Section 4reports the empirical estimation results, while Section 5 concludes.

**2. Challenges facing Saudi Arabia's vision 2030**

Saudi Arabia is the greatest regional power owing to its massive oil wealth, and also due to its new ambitions. The policy of wide-scale public works implemented by the government, as well as foreign direct investment and banking and financial soundness, have enabled Saudi Arabia to become the number one regional economy. Nevertheless, the economy of Saudi Arabia is entirely based on oil. This country has the world's second-largest proven petroleum reserves after Venezuela and it is the largest exporter of petroleum. Add to this, Saudi Arabia has the fifth-biggest proven natural gas reserves. Saudi Arabia is commonly regarded as an energy superpower. But since the 2014 oil price decline, the country is plagued by major economic hardships, which has forced it to reduce its public spending. Oil is still account for about 80 per cent of Saudi exports, and three-quarters of total tax revenue depend on it. The drop in oil prices since June 2014 created a certain obsession among Saudis with economic and political decline. The dramatic decline in oil prices has depleted Saudi Arabia's cash reserves by a whopping US150 billion dollars and pushed the ruling family to quickly contrive an effective rescue plan. Gigantic waves of change are sweeping across the Middle East region. The appointment of Prince Mohamed bin Salman as Crown Prince is part of this strategy. Previously it required the consent of the king's brothers and half-brothers of the king to pass on a project. Today, efficiency prevails. One should remember that the tradition in Saudi Arabia consisted of passing the 'Royal Scepter' among the sons of the kingdom founder, Ibn Saud, and not from father to son. This was a part of the internal politics driven by Ibn Saud many wives and dozens of children of. When Saudi Arabia's king Abdullah bin Abdul-Aziz died in January 2015 at the age of 90, the candidates for his replacement were no longer young men. Nevertheless, the transfer of the role to the next generation intensified anxiety of an internal civil war breaking out between many princes, a war that might have damaged the existence of the House of Saud. To deal with increasing fears, the successor was his half-brother Salman who enjoyed the entire confidence of the other brothers. When the brother designated as Crown Prince was very old (about 80) and with failing health, royal decisions would be lengthy preventing the system from functioning effectively; hence the mini-revolution that happened this year with the appointment of Prince MBS as Crown Prince. MBS was the sixth brother and the most talented among them. The achievement of a diversified economy is one of the main goals of the crown prince. MBS is taking the example of Abu Dhabi to develop its economy.

On April 25, 2016, Crown Prince Muhammad Bin Salman announced the "Vision 2030" plan to revolutionize the Saudi economy by reducing its heavy reliance on oil, diversifying its economy, and developing public service sectors including health, education, infrastructure and tourism. The main objectives of this plan incorporate reinforcing economic and investment activities and expanding non-oil industry trade among countries. The serious oil price collapse forced Saudi Arabia to undertake deeper changes to its economy. The Saudi government has imposed new taxes, including a 5 percent value added tax (Bouoiyour and Selmi 2019). It must be stressed that this is the first tax imposed in the country. The country has also accelerated its efforts to build a more diversified industrial economy, with new facilities for various sectors including chemicals, fertilizers, aluminum and cement. Regardless of MBS's reform efforts, shifting to a diversified economic structure seems difficult for Saudi Arabia. This is attributed to the fact that Saudi Arabia, as a "rentier state" and therefore, has had a limited incentive to spur the growth of any non-oil sector of its economy. It is also difficult to attract foreign investors when Saudi officials do not provide information about the volume of reserves of proven oil reserves. For boosting international investors' confidence and for Saudi Arabia's economic reforms to carry credibility, there is an urgent necessity for greater transparency in how government finances are generated and dispersed. Moreover, companies operating or planning to invest in Saudi Arabia face also a significant risk of corruption. The privatization of Saudi Aramco, which constitutes the barley point of this strategy of seduction, indefinitely postponed, according to Saudi sources. It is also difficult to attract foreign investors when Saudi officials do not provide information about the volume of reserves of proven oil reserves. Likewise, the company's accounts have never been audited. For boosting international investors' confidence and for Saudi Arabia's economic reforms to carry credibility, there is an urgent necessity for greater transparency in how government finances are generated and dispersed. According to the UN Conference on Trade and Development (UNCTAD, 2017)[1], inward investment into Riyadh dropped markedly in 2017, raising several questions regarding the prospects for the economic reform agenda being conducted by Crown Prince MBS.

---

[1] https://unctad.org/en/PublicationsLibrary/wir2018_overview_en.pdf

Recently, the case of the Saudi journalist, Jamal Khashoggi, killed inside the Saudi consulate in Istanbul hurted kingdom's image and deterred foreign investors to allocating to the country. This article' outcomes confirm that Jamal Khashoggi killing adversely affected Saudi Arabia stock market. The unexplained disappearance of Jamal Khashoggi has led big-name business partners to cut ties with projects that are of utmost importance to MBS's wishes of building a diversified and modern economy with technology at its core. In fact, the leader of the US bank JP Morgan and the co-chair of the board of US automaker Ford announced that he cancels his participation in the summit. Besides, Softbank CEO also announced to skip the event regardless of being a close ally of MBS ; similarly for the cases of Uber Technologies and America online (AOL) founders. In the same context, the German giant Siemens CEO as well as the leaders of the French banks BNP Paribas and Société Générale proclaimed that they monitor closely the situation. Likewise, the president of the World Bank, Jim Yong Kim, withdrew his participation. The New York Times, Financial Times, Bloomberg, CNN and CNBC also withdrew as media sponsors. Last year, New York Times claimed that Saudi Arabia is experiencing its own "Arab Spring", though it wasn't being prompted by tired and disappointed youth or angry protesters, but by the kingdom's powerful crown prince MBS. Less than 12 months later, the same American newspaper indicated that the Saudi journalist killing will be disastrous for Riyadh's diplomacy and the Saudi Arabia big economic ambitions.

3. **Methodology and data**

The event study methodology is designed to assess the impact of an event on the stock price of a firm. The investigation of such an effect consists of analyzing the changes in stock price beyond expectation (i.e., abnormal returns) during a specific period of time (i.e., event window). This paper adopts an event study methodology to examine the sectoral effects of Khashoggi killing on Saudi Arabia economic ambitions. We follow Ramiah et al. (2013, 2016) and Pham et al. (2018) by adjusting daily returns to obtain the ex-post-abnormal returns where adjustment is approximated by the Capital Asset Pricing Model (CAPM). We allow for the possible overreaction or under-reaction to the date of Jamal Khashoggi's disappearance whereby markets have a tendency to correct their mistakes in subsequent periods. We define day "0" as the announcement day of Khashoggi murder. Then, the estimation and event windows can be determined (Fig.1). The interval T0-T1 is the estimation window which provides the information needed to specify the normal return (i.e., prior to the occurrence of the event). The interval T2-T1+1 is the event window, and the interval T3-T2 is the post event window which is used to investigate the behavior of sectoral Saudi stock market following the event.

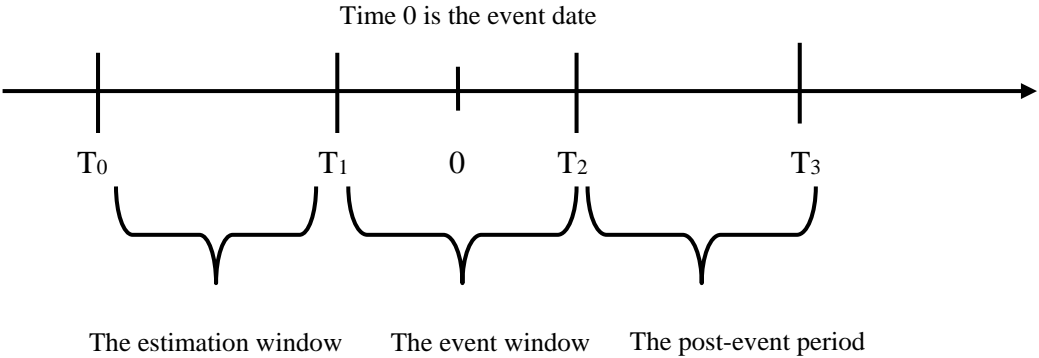

**Fig.1.** Data structure of an event study

Throughout this study, we estimate the cumulative abnormal returns over the following 2, 5 and 10 trading days to show whether or not the market reverts back to its mean process or continues to deviate from the mean price. The cumulative abnormal return (CAR) for a sector i during the event window $[\tau_1; \tau_2]$ surrounding the event day t = 0, where $[\tau_1; \tau_2] = \in [-2;+2]$, $[-5;+5]$ and $[-10;+10]$. Then and to evaluate the immediate change in systematic risk, we adjust the CAPM by including an interaction variable. Our immediate risk model detects the average change in risk resulting from Khashoggi killing. A dummy variable (DV), which takes the value of one on the first day of trading after the Jamal's disappearance and zero otherwise, is created to depict the immediate changes in systematic risk. This DV is multiplied by the market risk premium to form the interaction variable. Based on Ramiah et al. (2013)'s study, the model to be estimated is denoted as

$$\tilde{r}_{it} - \tilde{r}_{ft} = \beta_i^0 + \beta_i^1 [\tilde{r}_{mt} - \tilde{r}_{ft}] + \beta_i^2 [\tilde{r}_{mt} - \tilde{r}_{ft}] * DV + \beta_i^3 DV_t + \tilde{\varepsilon}_{it} \quad (1)$$

where $\tilde{r}_{it}$ is industry i's return at time t, $\tilde{r}_{ft}$ is the risk free rate at time t, $\tilde{r}_{mt}$ is the market return at time t, DV is a DV that takes the value of one on the first day of trading following the the disappearance of Jamal Khashoggi on 2 october in Saudi consulate in Istanbul and zero otherwise, $\beta_i^0$ is the intercept of the regression equation [E($\beta_i^0$) = 0], $\beta_i^1$ corresponds to the average short-term systematic risk of the industry, $\beta_i^2$ denotes the change in the industry risk, and $\beta_i^3$ corresponds to the change in the intercept, $\tilde{\varepsilon}_{it}$ is the error term,. The Equation (1) is estimated to determine the short-term change in systematic risk of the Saudi disaggregated stock markets.

The event study methodology has been largely criticized for several reasons including the non-normality of the abnormal return distribution and the effect of stock market synchronization. This study proposes improvements to avoid these shortcomings. Specifically, we use the Corrado (1989) non-parametric ranking test and the non-parametric conditional distribution suggested by Chesney et al. (2011) to study the repercussions of the uncertainty surrounding Khashaoggi murder. To make certain that the emerging abnormal returns are due entirely to the impact of the Saudi journalist killing, we account for possible asynchronicity, market integration and spillover effects by controlling for market risk premia (i.e., the difference between the expected market return and the risk-free rate) representing Asia, Europe and the United States.

To determine the immediate risk associated with Khashoggi crisis, we adjust the CAPM by including interaction variables. This enables to capture the average change in risk as a consequence of Jamal Khashaoggi's death. A dummy variable which takes the value of one on the first day of trading after the date of the journalist disappearance and zero otherwise, is created to detect the immediate changes in systematic risk. We multiplied this dummy variable by the market risk premium to form the interaction variable. Daily disaggregated stock price data series over the period 21 June 2017– 01 December 2018 were extracted from the Thomson Reuters database. The selected industries include Financials (banks, insurance, reinsurance and financial services), Materials, Oil and Gas (oil and gas producers, oil equipment, and services, distribution and alternative energy) and Technology (software and computer services, and technology hardware and equipment). Each sector index represents a capitalization-weighted portfolio of the largest Saudi companies in this sector. We transformed all the variables by taking natural logarithms to correct for heteroskedasticity and dimensional differences.

4. **Empirical results**

   *4.1. Main findings*

Our hypotheses are based on discussions pertaining to possible detrimental consequences of Khashoggi crisis on Saudi Arabia economy. For instance, a discussion in CNN Business leads us to believe that the Saudi Journalist murder would draken the kingdom's big economic dreams. Before Khashoggi killing, more than 3,500 business leaders and government officials have confirmed their attendance for the event 'Davos in the Desert'. MBS promised global CEOs and international investors that the government will create a transparent, secure, stable and understandable business climate. But the disappearance of Jamal Khashoggi deter businesses and CEOs from investing in Riyadh, harming the kingdom's bid to transform its oil-dependent economy. We utilize this argument to formulate the hypothesis that khashoggi killing is bad news for the Saudi Arabia business, leading to negative ARs and higher risk for different sectors.

Our results confirm the hypothesis that the outcome of the Khashoggi had an adverse impact on stock returns. Indeed, the majority of sectors exhibit negative abnormal returns (ARs), but we also document positive ARs for oil and gas sector. Tables 1 and 2 display a summary of the estimated ARs. It is noticeable that Financials, Materials and Technology were influenced the most. The results of risk analysis (reported in Table 3) produce evidence of changes in short-term systematic risk. There seems to be an industry effect whereby the impact of Khashoggi killing on industry betas varies sharply across the industries under study.

Table 1 summarizes ARs and CARs (2, 5 and 10 days) and their t-statistics following the disappearance of Jamal Khashoggi. We show that the Saudi stock market reacted negatively to this event. This holds true for most sectors. Banking and financial services were negatively affected by -5.09% after 2 days and -9.16 % after 10 days. Materials were adversely influenced by -2.74% on the first day of trading, and by about -8.6% after 10 days. Software and computer services (i.e., technology) were negatively impacted by −3.64% after 5 days and by -4.52 after 10 days. The Oil and Gas sector responds differently to this event. It reacts positively by 1.49 on the first day of trading, negatively and modestly by -0.37% after 2 days, while it did not record any statistically significant results up to 10 days.

**Table 1.** Sectoral reactions in Saudi Arabia following Khashoggi killing

| Sectors | AR | t-Stat | CAR2 | t-Stat | CAR5 | t-Stat | CAR10 | t-Stat |
|---|---|---|---|---|---|---|---|---|
| Financials | -4.36*** | -3.91 | -5.09*** | -4.86 | -8.07*** | -5.43 | -9.16*** | -4.89 |
| Materials | -2.74*** | -4.62 | -6.41*** | -7.56 | -7.11* | -1.91 | -8.59*** | -5.21 |
| Technology | -3.12*** | -5.71 | -4.36** | -2.61 | -3.64* | -1.75 | -4.52** | -2.79 |
| Oil and Gas | 1.49** | 2.89 | -0.37* | -1.92 | 1.01 | 1.13 | -0.08 | -0.63 |

Notes: AR: Abnormal returns; CAR: Cumulative abnormal returns; ∗, ∗∗, ∗∗∗ denote statistical significance at the 10%, 5% and 1% levels, respectively.

Fig. 2 graphically describes the cumulative abnormal return performance of Saudi Arabian companies prior to and following Jamal Khashoggi killing on 2 October. As mentioned above, positive and negative CARs imply favorable and unfavorable outcomes, respectively. We show that the Saudi stock price responses of different sectors surrounding the Khashoggi murder appear dissimilar either for the announcement disappearance day CAR or the [−5; + 5] event window CAR. We clearly observe that the Saudi journalist's death is associated to severe stock prices declines for Financials, Materials and Technology from the day relative to the announcement of the journalist's disappearance (t=0). However, oil and gas seems less sensitive to Khashoggi killing.

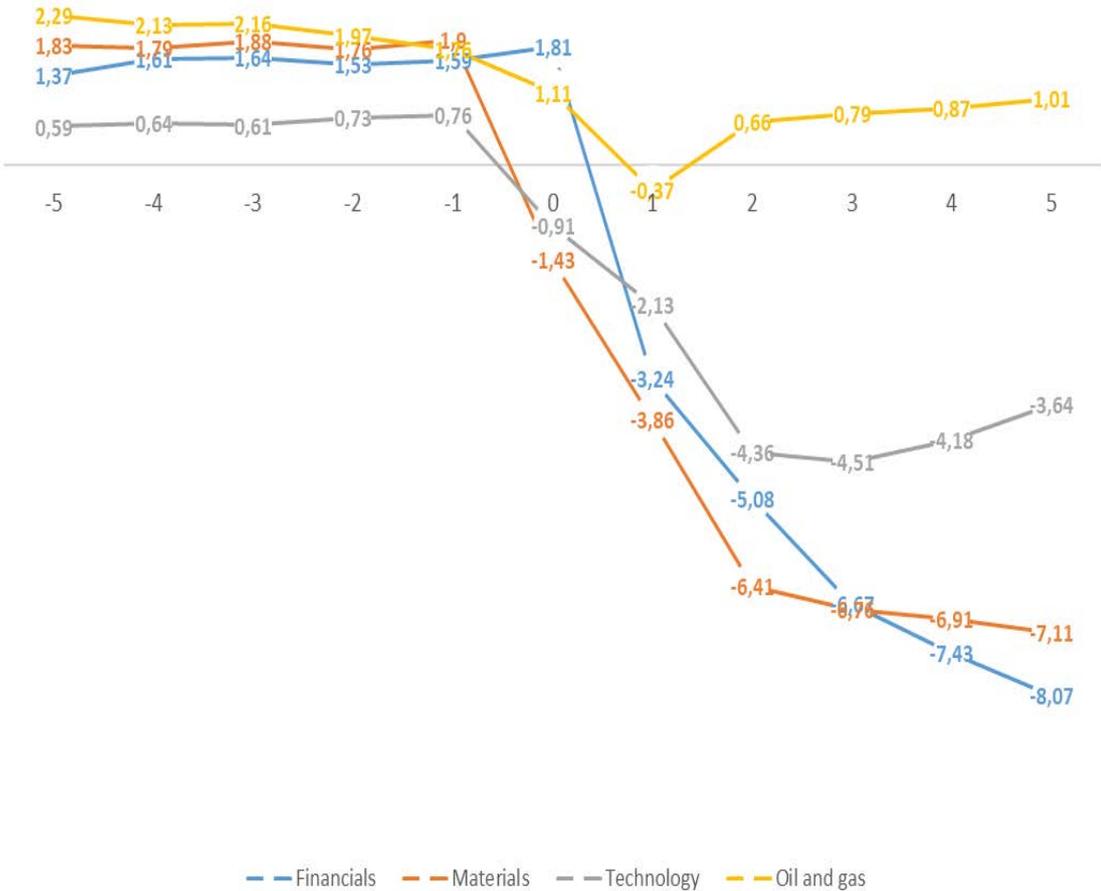

**Fig. 2.** The cumulative abnormal returns of Saudi Arabia stock price index in response to Khashoggi killing by industry: [−5; + 5] event window

We document that these findings are supported by the robustness test results displayed in Table 2. After controlling for asynchronicity and market integration, we consistently find that the khashoggi killing had harmfully affected banks and financial services, materials, and technology. The effect on Oil and Gas industry, nevertheless, was relatively weak or insignificant. Likewise, the results derived from the non-parametric conditional probability of Chesney et al. (2011) confirm that Financials, Materials and Technology were the most harmfully affected by Khashoggi murder.

**Table 2.** Robustness tests for sectoral responses in Saudi Arabia following Khashoggi killing

| Sectors | Conditional probability | | | Market integration | |
|---|---|---|---|---|---|
| | $t_{Corrado}$ | CP | t-stat | AR | t-Stat |
| Financials | -5.14 | 0.56 | 2.59 | -8.55** | -2.76 |
| Materials | -4.39 | 0.43 | 1.86 | -6.91*** | -5.10 |
| Technology | -4.11 | 0.29 | 3.14 | -5.62* | -1.91 |
| Oil and Gas | 0.17 | 0.02 | 1.21 | 0.11 | 0.74 |

Notes: $t_{Corrado}$: The Corrado (1989) non-parametric ranking test; CP: The non-parametric conditional probability proposedby Chesney et al. (2011); AR: Abnormal returns; ∗, ∗∗, ∗∗∗ denote statistical significance at the 10%, 5% and 1% levels, respectively.

Interestingly, the changes in the short-term systematic risk following Khashoggi murder by sector are reported in Table 3. The results reveal that Khashoggi crisis has prompted a sharp increase in systematic risk. The increase in systematic risk by moving from the period prior to Jamal's disappearance to the post-Khashoggi killing period is valid for Financials, Materials and Technology. It is also worth mentioning that the Oil and Gas industry did not witness any change in systematic risk.

**Table 3.** Changes in short-term systematic risk of the Saudi Arabia stock market sectors following Khashoggi killing

| | Beta prior to Khashoggi killing | Immediate risk | Beta post- Khashoggi killing |
|---|---|---|---|
| Financials | 0.46 | 0.71 | 0.92 |
| Materials | 0.52 | 0.69 | 0.88 |
| Technology | 0.33 | 0.46 | 0.57 |
| Oil and Gas | 0.31 | 0.36 | 0.35 |

To further ascertain the robustness of our results, we perform a variety of econometric tests on all of the regression models. The Chow test is employed to capture the occurrence or otherwise of structural breaks following the journalist murder, the Wald test is utilized to control for redundant variables, AR and MA terms are incorporated to account for possible autocorrelation and different GARCH specifications (symmetric versus asymmetric and linear versus nonlinear) are used to correct for the ARCH effects. The findings drawn from these tests are available for interested readers upon request.

*4.2. Discussion of results*

This research adopts an event study methodology to examine the implications of Jamal Khashoggi killing for Saudi Arabia business, as measured by abnormal returns. We find the death of Saudi Journalist has a mixed effect on abnormal returns with clear sector-by-sector differences. The results reveal that the banks and financial services, materials, and technology were affected negatively, whereas oil and gas sector was not significantly influenced.

The fact that oil industry appears unharmed by the gruesome murder of Jamal Khashoggi, is not surprising. The oil price dynamics are the result of complex processes happening within the global economy. There is no single systematic explanation for oil price movements. Although the market powers of oil producers and the strategic position of OPEC members within the group of oil producing countries (in particular, Saudi Arabia) play a potential role in explaining oil price variation, other factors (supply, demand, speculation and geopolitics) contribute significantly to its fluctuations (Breitenfellner et al. 2009). In addition, the spare production capacity has a vital role in the world oil market, and the spare capacity held by core OPEC members (i.e., Saudi Arabia, Kuwait and the UAE) especially is greatly significant for its ability to offset demand and supply shocks and stabilize the oil market.[2]

---

[2] The OPEC members have engaged to balance the market by adjusting their production to counteract oil demand and supply shocks.

While the OPEC core countries do not act as an homogeneous unit to effectively manage spare capacity, each member seems to have taken earnestly the responsibility to calm oil market. In any unforeseen event, the volume of spare capacity held by the OPEC core countries is the largest (Pierru et al. 2018). Moreover, crude oil futures enable to form expectations with respect future prices, and these expectations may be considered as a determinant of oil price. Indeed, oil future contracts enable market participants to look in today a price at which to buy or sell a fixed quantity of the commodity on a specific date in the future. These financial instruments allow traders to minimize individual default risk, making the futures market a mechanism for hedging and/or speculating on oil price risks (Kilian and Murphy 2014).

The negative responses of banks and financial services, materials, and technology sectors can be attributed to the fact that foreign investors lose their confidence and pull money out of the kingdom. The announcement of the classification of Saudi Arabia as an emerging market category in June 2018come along with significant risks. First, the significant impact of the difference in trading days between Saudi Arabia stock market and other major markets which could exacerbate the short-run volatility. Saudi capital markets are operational from Sunday to Thursday and are closed on Friday and Saturday. Most economic and business information in the world is released during the week from Monday to Friday. Such dissimilarity in trading days could also lead to high volatility in the Saudi Arabian stock market. Second and more interestingly, the wide entry of international investments into the Saudi stock market could yield to an increased movement in short-term capital. In other words, as part of MSCI emerging market, it was highly expected that this status would ease inflows of foreign money into the Saudi economy. However, with Jamal Khashoggi's murder, international investors pull back from country's share market. These large scale capital outflows would lead to excessive fluctuations in the stock prices.

## 5. Conclusions

Despite the international praise for Mohamed Bin Salman (popularly known as MBS)'s large-ranging plan for social and economic reforms, known as 'Vision 2030', human rights groups have voiced concerns over war crimes committed in Yemen, the Qatar diplomatic crisis, the tensions between Saudi Arabia and Lebanon following the mysterious resignation of prime minister, the arrest of members of the Saudi royal family and lastly the death of Jamal Khashoggi. Although the crown prince's public image was unharmed by the first events, Khashoggi killing has considerably changed the narrative on MBS. This study applies a variety of methods (i.e., an event-study methodology, an adjusted asset pricing model, a Corrado ranking test, and a non-parametric conditional probability) to examine the responses of various Saudi Arabian industries to the murder of Jamal Khashoggi at the kingdom's consulate in Istanbul on 2 October.

The results presented throughout this research reveal that Khashoggi killing is a bad news for Saudi Arabia's ambitious social and economic plan. With the exception of Oil and Gas, all the sectors (in particular, banks and financial services, materials, and technology) responded harmfully to this event as indicated by negative ARs and CARs. Also, an increase in short-term systematic risk of most Saudi Arabia stock market sectors following Khashoggi's disappearance.

Saudi Arabia's drastic efforts to modernize its stock market and attract international investments were recognized by MSCI emerging markets index. By supporting the incorporation of Saudi Arabia in emerging markets, international institutional investors are able to access and operate in this market. Our results clearly underscore that foreign investors were rattled by Riyadh's noticeable deteriorating links with foreign governments following the death of journalist Jamal Khashoggi. The analysis of Khashoggi case highlights that why need businesses is the assurance that the political environment is stable enough, and the critical thinking and the freedom to question are encouraged. A recent literature documented that multinational corporations benefit from improved freedom of expression in the countries where they operate. Arguably, Masrorkhah and Lehnert (2017) claimed that stock markets in countries with a free press are likely to be better processors of economic information. This can be advanced as an element of explanation for the adverse reactions of Saudi stocks to Khashoggi fallout. It must be added that corruption has long been endemic in Saudi Arabia. According to the 2017 Corruption Perceptions Index, corruption rank in Saudi Arabia averaged 62.47 from 2003 until 2017. The Russian experience has demonstrated that international investors are willing to provide funds and the required managing expertise to privatized companies only if the legal and political infrastructure is appropriate, aiming at restraining corruption among government officials and mitigating the heightened risks of expropriation (Lombardo and Pagano, 2002).